\documentclass[journal,10pt,twocolumn]{IEEEtran}

\usepackage{cite}
\usepackage[T1]{fontenc}
\usepackage{graphicx}
\usepackage{amsmath}
\usepackage{amssymb}
\usepackage{multirow}

\newcommand{\secref}[1]{\mbox{Section~\ref{#1}}}
\newcommand{\figref}[1]{\mbox{Fig.~\ref{#1}}}

\renewcommand{\Pr}[1]{\ensuremath{\mathrm{P}\!\left[#1\right]}}

\DeclareMathOperator*{\argmin}{arg\;min}
\DeclareMathOperator*{\argmax}{arg\;max}

\DeclareMathOperator*{\define}{\triangleq} 

\newcommand{\abs}[1]{\ensuremath{\left|#1\right|}}

\newcommand{\C}{\ensuremath{\mathbb{C}}}
\newcommand{\R}{\ensuremath{\mathbb{R}}}
\newcommand{\Z}{\ensuremath{\mathbb{Z}}}
\newcommand{\CZ}{\ensuremath{\mathbb{CZ}}}

\newcommand{\setD}{\ensuremath{\mathcal{D}}}

\newcommand{\setL}{\ensuremath{\mathcal{L}}}

\newcommand{\setO}{\ensuremath{\mathcal{O}}}

\newcommand{\setX}{\ensuremath{\mathcal{X}}}

\newcommand{\bmb}{\ensuremath{\mathbf{b}}}
\newcommand{\bmc}{\ensuremath{\mathbf{c}}}

\newcommand{\bmn}{\ensuremath{\mathbf{n}}}

\newcommand{\bmr}{\ensuremath{\mathbf{r}}}
\newcommand{\bms}{\ensuremath{\mathbf{s}}}

\newcommand{\bmu}{\ensuremath{\mathbf{u}}}
\newcommand{\bmv}{\ensuremath{\mathbf{v}}}
\newcommand{\bmw}{\ensuremath{\mathbf{w}}}
\newcommand{\bmx}{\ensuremath{\mathbf{x}}}
\newcommand{\bmy}{\ensuremath{\mathbf{y}}}

\newcommand{\bmshat}{\ensuremath{\hat{\bms}}}

\newcommand{\bmuhat}{\ensuremath{\hat{\bmu}}}

\newcommand{\bmxhat}{\ensuremath{\hat{\bmx}}}

\newcommand{\bmntilde}{\ensuremath{\tilde{\bmn}}}

\newcommand{\bmrtilde}{\ensuremath{\tilde{\bmr}}}

\newcommand{\bA}{\ensuremath{\mathbf{A}}}
\newcommand{\bB}{\ensuremath{\mathbf{B}}}

\newcommand{\bG}{\ensuremath{\mathbf{G}}}
\newcommand{\bH}{\ensuremath{\mathbf{H}}}
\newcommand{\bI}{\ensuremath{\mathbf{I}}}

\newcommand{\bQ}{\ensuremath{\mathbf{Q}}}
\newcommand{\bR}{\ensuremath{\mathbf{R}}}

\newcommand{\bT}{\ensuremath{\mathbf{T}}}

\newcommand{\bOne}{\ensuremath{\mathbf{1}}}

\newcommand{\MT}{{\ensuremath{M}}}
\newcommand{\MR}{{\ensuremath{M}}}
\newcommand{\SNR}{\ensuremath{\mathrm{SNR}}}

\newcommand{\ML}{{\ensuremath{\mathrm{ML}}}}

\title{Finite Lattice-Size Effects in MIMO Detection}

\author{ 
  C.~Studer\IEEEauthorrefmark{1},
  D.~Seethaler\IEEEauthorrefmark{3}, and
  H.~B\"olcskei\IEEEauthorrefmark{3} \vspace{0.5cm} \\
\begin{minipage}[t]{0.46\textwidth}
  \centering
  \IEEEauthorrefmark{1}Integrated Systems Laboratory \\
  ETH Zurich, 8092 Zurich, Switzerland \\
  e-mail: studer@iis.ee.ethz.ch
\end{minipage}
\begin{minipage}[t]{0.46\textwidth}
  \centering
  \IEEEauthorrefmark{3}Communication Technology Laboratory \\
  ETH Zurich, 8092 Zurich, Switzerland \\
  e-mail: \{seethal,boelcskei\}@nari.ee.ethz.ch
\end{minipage}
\thanks{This work was supported in part by the STREP project
  No.~IST-026905 (MASCOT) within the Sixth Framework Programme (FP6)
  of the European Commission.}\vspace{-0.0cm}}

\newcommand{\figspacepre}{\vspace{-0.3cm}} 
\newcommand{\figspacepost}{\vspace{-0.5cm}} 
\newcommand{\figscale}{0.75} 

\newcommand{\DREL}{{\ensuremath{\underline{D}}}}
\newcommand{\CREL}{{\ensuremath{\underline{C}}}}

\newcommand{\bmxREL}{{\ensuremath{\underline{\mathbf{x}}}}} 
\newcommand{\bmuREL}{{\ensuremath{\underline{\mathbf{u}}}}}
\newcommand{\bmuhatREL}{{\ensuremath{\hat{\underline{\mathbf{u}}}}}} 
\newcommand{\bmxhatREL}{{\ensuremath{\hat{\underline{\mathbf{x}}}}}} 
\newcommand{\CZMT}{{\ensuremath{(\CZ)^{\MT}}}} 

\newcommand{\LatticeREL}{{\ensuremath{\underline{\setL}}}} 
\newcommand{\Lattice}{{\ensuremath{\setL}}}

\newcommand{\BABAI}{{\ensuremath{\mathrm{B}}}}

\begin{document}

\maketitle

\begin{abstract}
  Many powerful data detection algorithms employed in multiple-input
  multiple-output (MIMO) communication systems, such as sphere
  decoding (SD) and lattice-reduction (LR)-aided detection, were
  initially designed for infinite lattices.  Detection in MIMO systems
  is, however, based on finite lattices. In this paper, we
  systematically study the consequences of finite lattice-size for the
  performance and complexity of MIMO detection algorithms formulated
  for infinite lattices. Specifically, we find, considering
  performance and complexity, that LR does not seem to offer
  advantages when used in conjunction with~SD.
\end{abstract}


\section{Introduction} \label{introduction}

The computational complexity associated with maximum-likelihood (ML)
detection (MLD) in multiple-input multiple-output (MIMO) communication
systems poses significant challenges for corresponding VLSI
implementations~\cite{burg05}. Promising approaches for complexity
reduction make explicit use of the structure in the problem by
transforming it into an equivalent lattice decoding\footnote{In the
  context of this paper it would be more appropriate to use the term
  ``lattice detection''.}  problem, where the underlying lattice is
finite. Some of the most prominent techniques for efficient lattice
decoding were, however, developed for infinite lattices, e.g., sphere
decoding~(SD)~\cite{FP1985} or lattice-reduction (LR)-aided detection,
e.g.,~\cite{YW02,WF03}.

The goal of this paper is to investigate the fundamental differences,
in terms of complexity and performance, between MIMO detection on
finite and on infinite lattices. We show that relaxation of the MIMO
detection problem to infinite lattices followed by LR can reduce the
complexity of SD at the cost of a significant performance loss in
terms of error rate.  This performance loss is caused by symbol
estimates that do not belong to the finite lattice and hence require
remapping (onto the finite lattice). We demonstrate that (near-)ML
achieving remapping strategies entail a complexity which is comparable
to that of SD operating directly on the finite lattice, thus rendering
LR in conjunction with SD unattractive for practical applications.

\subsubsection*{Notation}

Matrices are set in boldface capital letters, vectors in boldface
lowercase letters. The superscripts~$^T$ and~$^H$ stand for transpose
and conjugate transpose, respectively. We write~$A_{i,j}$ for the
entry in the~$i$th row and~$j$th column of the matrix~$\bA$,~$b_i$ for
the~$i$th entry of the
vector~\mbox{$\bmb=[\,b_1\,\,b_2\,\,\cdots\,\,b_N\,]^T$}, and we
denote the~\mbox{$\ell^2$-norm} of \mbox{$\bmb\in\C^N$} as~$\|\bmb\|$.
$\bI_N$ stands for the $N\times N$ identity matrix and $\bOne_{N}$
denotes the $N$-dimensional all-ones vector. $|\setO|$ designates the
cardinality of the set~$\setO$ and~$\CZ$ stands for the set of
Gaussian integers, i.e.,~$\CZ = \Z +\sqrt{-1}\,\Z$.  The real and
imaginary part of~$x\in\C$ is denoted by $\Re\{x\}$ and $\Im\{x\}$,
respectively.


\section{MIMO Detection as Lattice Decoding}
\label{latticedetection}

Consider a coherent MIMO system, i.e., the receiver has perfect
channel state information (CSI) while the transmitter does not have
CSI, with~$\MT$ transmit and~$\MR$ receive antennas\footnote{The
  restriction to the number of transmit antennas being equal to the
  number of receive antennas is made for simplicity of exposition. The
  results are also valid for the number of receive antennas being
  larger than the number of transmit antennas.}. The bit-stream to be
transmitted is mapped to $\MT$-dimensional transmit symbol
vectors~$\bms\in\setO^\MT$, where~$\setO$ corresponds to the
underlying scalar QAM constellation. The associated complex baseband
input-output relation is given by
\begin{align} \label{eq:MIMOchannelmodel}
\bmy = \bH\bms + \bmn
\end{align}
where~$\bH$ stands for the $\MR\times\MT$ channel matrix,~$\bmy$ is
the~$\MR$-dimensional received signal vector, and~$\bmn$ is an i.i.d.\
circularly symmetric complex Gaussian distributed noise vector of
dimension~$\MR$ with variance $N_o$ per complex entry. The
signal-to-noise ratio is defined as~$\SNR=E_s/N_o$, where~$E_s$
denotes the average signal power per receive antenna. 

\subsection{Transformation to Lattices}

In order to see that the MIMO detection problem is equivalent to a
lattice decoding problem, we start by mapping the elements~$s\in\setO$
to elements~$x\in\CZ$ using the transformation~$x = as+c$.  The
constants~$a,c\in\R$ with~$a>0$ and~$c>0$ are independent of~$s$ and
are chosen such that~$x\in\setX\subset\CZ$
with~$\abs{\setX}=\abs{\setO}$ and
\begin{align} \label{eq:formalsetdefinition} \setX = \big\{ x\in\CZ
  \mid & (k_\mathrm{min} \leq \Re\{x\} \leq k_\mathrm{max}) \nonumber
  \\ & \land
  (k_\mathrm{min} \leq \Im\{x\} \leq k_\mathrm{max}) \big\}
\end{align}
where~$k_\mathrm{min},k_\mathrm{max}\in \Z$. Note
that~\eqref{eq:formalsetdefinition} can be used for square QAM
constellations to check whether~$x'\in\CZ$ is in~$\setX$ by performing
separate boundary checks for the real and imaginary part of~$x'$. In
the case of non-square QAM constellations the boundary checks take a
slightly more complicated form. The transmit
vectors~$\bms\in\setO^\MT$ can be mapped to
vectors~$\bmx\in\setX^\MT\subset\CZMT$ according to
\begin{align} \label{eq:constellationtrafo}
  \bmx=a\bms+\bmc
\end{align}
where~$\bmc=c\bOne_{\MT}$. The inverse transformation associated
with~\eqref{eq:constellationtrafo} is given by~$\bms =
a^{-1}(\bmx-\bmc)$. The input-output
relation~\eqref{eq:MIMOchannelmodel} can now be transformed into
\begin{align} \label{eq:MIMOlatticechannelmodel}
\bmr = \bG\bmx + \bmn
\end{align}
where~$\bG=a^{-1}\bH$ and~$\bmr=\bmy+\bG\bmc$ is a translated version
of the received vector~$\bmy$.  The essence of the transformation
of~\eqref{eq:MIMOchannelmodel} into~\eqref{eq:MIMOlatticechannelmodel}
is that now the received vector~$\bmr$ can be interpreted as a lattice
point~$\bmu\in\Lattice(\bG)$ that has been translated by the additive
Gaussian noise vector~$\bmn$. Here,
\begin{align} \label{eq:finitelattice} 
  \Lattice(\bG) \define \Big\{ \bG\bmx \mid \bmx\in\setX^\MT \Big\}
\end{align}
denotes the finite lattice generated by~$\bG$. In the remainder of the
paper, we shall work with the input-output
relation~\eqref{eq:MIMOlatticechannelmodel} exclusively.

\subsection{ML Detection}

MLD in MIMO systems computes the estimate
\begin{align} \label{eq:constrainedMLrule}
  \bmuhat^\mathrm{ML} = D^\ML(\bmr) = \argmin_{\bmu\in\Lattice(\bG)}\|\bmr-\bmu\|
\end{align}
which amounts to solving a closest-vector problem (CVP) in the finite
lattice~$\Lattice(\bG)$. Since each lattice point in~$\Lattice(\bG)$
is associated with a transmit vector in~$\setX^\MT$ according to the
relation~\mbox{$\bmu=\bG\bmx$}, the ML-estimate~$\bmuhat^\ML$ obtained
by solving~\eqref{eq:constrainedMLrule} can be transformed
into\footnote{Note that we implicitly assume that~$\bG$ has full
  rank; this is satisfied, for example, with probability 1, if the
  entries of~$\bG$ are i.i.d.\ circularly symmetric complex Gaussian
  distributed.  Furthermore, in practice~$\bG^{-1}$ does not have
  to be computed explicitly as~$\bmu$ in~\eqref{eq:constrainedMLrule}
  can be replaced by~$\bG\bmx$ and the minimization can be performed
  over~$\bmx\in\setX^\MT$.} \mbox{$\bmxhat^\ML=\bG^{-1}\bmuhat^\ML$}, which upon
inversion of~\eqref{eq:constellationtrafo} yields
the ML-estimate~$\bmshat^\ML$.  Solving~\eqref{eq:constrainedMLrule}
through an exhaustive search over all lattice points
in~$\Lattice(\bG)$ typically results in high computational complexity
as $\abs{\Lattice(\bG)}=\abs{\setX}^\MT$ may be large.

\subsection{Relaxation and Lattice Reduction}
\label{relaxationandLR}

A promising approach to reducing the computational complexity
associated with solving~\eqref{eq:constrainedMLrule} is to relax the
finite lattice in~\eqref{eq:finitelattice} to the infinite
lattice\footnote{In the remainder of the paper, underlined quantities
  always refer to the infinite-lattice case.}
\begin{align} \label{eq:infinitelattice} 
  \LatticeREL(\bG) \define \Big\{ \bG\bmxREL \mid
  \bmxREL\in\CZMT \Big\}
\end{align}
and to solve a relaxed version of~\eqref{eq:constrainedMLrule} by
computing
 \begin{align} \label{eq:unconstrainedMLrule} \bmuhatREL^\ML =
    \DREL^\ML(\bmr) =
    \argmin_{\bmuREL\in\LatticeREL(\bG)}\|\bmr-\bmuREL\|
\end{align}
i.e., by searching over the infinite lattice generated by~$\bG$. This
search can be carried out more efficiently by using algorithms from
lattice theory, e.g., SD~\cite{FP1985} possibly in conjunction with LR
through the LLL algorithm~\cite{LLL82}. As the resulting
estimate~$\bmuhatREL^\ML$ is not necessarily in~$\Lattice(\bG)$,
remapping of~$\bmuhatREL^\ML$ onto the finite lattice~$\Lattice(\bG)$
is required whenever \mbox{$\bmuhatREL^\ML\notin\Lattice(\bG)$}.  In
the remainder of the paper, the term ``MLD'' always refers to the
finite-lattice MLD problem~\eqref{eq:constrainedMLrule}, whereas
``relaxed MLD'' is used to refer to the infinite-lattice MLD
problem~\eqref{eq:unconstrainedMLrule}.

The main motivation for relaxation resides in the fact that LR can be
applied only to infinite lattices. LR computes an equivalent and
``more orthogonal'' basis (for the lattice~$\LatticeREL(\bG)$) with
the generator matrix \mbox{$\bB=\bG\bT$}, where~$\bT$ is an
$\MT\times\MT$ unimodular matrix, i.e., $\abs{\det(\bT)}=1$
with~$T_{i,j}\in\CZ$ ($\forall i,j$). Thanks to the unimodularity
of~$\bT$, we have \mbox{$\LatticeREL(\bB)=\LatticeREL(\bG)$}.  It is
important to note that this equivalence holds only for infinite
lattices, in contrast to finite lattices where \mbox{$\Lattice(\bB)
  \neq \Lattice(\bG)$}, in general.  We therefore conclude that
relaxation to an infinite lattice is essential for the application of
LR to MIMO detection.

We emphasize that LR, in general, reduces the complexity of a
subsequent SD step and results in (often significant) performance
improvements when followed by sub-optimal detectors. In particular,
in~\cite{TMK07} it is shown that LR followed by linear detection
achieves full diversity.

\subsection{Schnorr-Euchner Sphere Decoding}
\label{spheredecoder}

SD was initially designed for the efficient solution of CVPs on
infinite lattices~\cite{FP1985}, but has also proven very efficient
for MLD~\cite{Mow91,VB93}. In the following, we briefly review the
main ingredients of Schnorr-Euchner SD (SESD) with radius reduction
for the finite-lattice case~\cite{AEVZ02}.  

The algorithm starts by performing a QR decomposition (QRD) of~$\bG$
according to~$\bG=\bQ\bR$, where the $\MR\times\MT$ matrix~$\bQ$ is
unitary and the upper-triangular~$\MT\times\MT$ matrix~$\bR$ has
real-valued non-negative entries on its main diagonal.
Left-multiplying~\eqref{eq:MIMOlatticechannelmodel} by~$\bQ^H$ leads
to the modified input-output relation $\bmrtilde = \bR\bmx +
\bmntilde$, where $\bmrtilde=\bQ^H\bmr$ and $\bmntilde=\bQ^H\bmn$ is
again (thanks to~$\bQ$ being unitary) i.i.d.\ circularly symmetric
complex Gaussian distributed with variance~$N_o$ per complex entry.
We arrange the partial symbol vectors (PSVs)
\mbox{$\bmx^{(i)}=[\,x_i\,\,x_{i+1}\,\,\cdots\,\,x_\MT\,]^T$} in a
tree that has its root just above level~$i=\MT$ and leaves on
level~$i=1$; the leaves correspond to vectors~$\bmx\in\setX^\MT$.
Each PSV is associated with a partial distance~(PD) according to
\begin{align} \label{eq:partialdistance} 
  d\big(\bmx^{(i)}\big) = \big\|\bmrtilde^{(i)}-\bR_i\bmx^{(i)}\big\|,
  \quad i=1,2,\ldots,\MT
\end{align}
where~$\bmrtilde^{(i)}=[\,\tilde{r}_i\,\,\tilde{r}_{i+1}\,\,\cdots\,\,\tilde{r}_\MT\,]^T$
and~$\bR_i$ contains the lower-right $(\MT-i+1)\times
(\MT-i+1)$-dimensional sub-block of~$\bR$. The MLD problem has
therefore been transformed into a weighted tree-search problem, where
PSVs and PDs are associated with nodes in the tree. For brevity, we
shall often say ``the node~$\bmx^{(i)}$'' meaning the node
corresponding to the PSV~$\bmx^{(i)}$. Each path through the tree
(from the root node down to a leaf node) corresponds to a symbol
vector~\mbox{$\bmx\in\setX^{\MT}$}.  Note that in the infinite-lattice
case, the tree-search problem needs to be relaxed to
vectors~$\bmxREL\in(\CZ)^\MT$.  The solution
of~\eqref{eq:constrainedMLrule} is given by the leaf associated with
the smallest metric.

The essence of SD is to constrain the search in the tree through a
sphere constraint (SC), realized by searching only those nodes that
satisfy~$d\big(\bmx^{(i)}\big)\leq r$, where~$r$ is the search radius
associated with a hypersphere centered in the received point.  SESD
refers to conducting the tree search in a depth-first manner and
visiting the children of a given node in ascending order of their
PDs~\cite{AEVZ02}. Since the PDs~$d\big(\bmx^{(i)}\big)$ are
non-decreasing as a function of~$i=\MT,\MT-1,\ldots,1$, a
node~$\bmx^{(i)}$ violating the SC can be pruned along with the entire
subtree emanating from that node. The algorithm initializes~$r=\infty$
and performs the update~$r\gets d(\bmx)$ whenever a valid leaf
node~$\bmx$ has been found. This approach is known in the literature
as radius reduction~\cite{AEVZ02,burg05}.

The complexity measure employed in this paper is given by the number
of nodes visited by SESD in the tree search, including the leaf nodes,
but excluding the root node. This complexity measure is directly
related to the VLSI implementation complexity for the finite-lattice
case\footnote{We expect this complexity measure to be equally relevant
  for the infinite-lattice case.}~\cite{burg05}.  The complexity
required for preprocessing (i.e., QRD or LR) will be ignored
throughout the paper. We note, however, that in latency-critical
applications, the preprocessing complexity can be critical.

In the remainder of the paper, ``SESD'', ``relaxed SESD'', and
``LR-aided SESD'' refer to SESD on finite lattices, on infinite
lattices, and to LR followed by SESD on infinite lattices,
respectively.


\section{Complexity of SESD and of Relaxed SESD}
\label{detectioncomplexity}

We next analyze the differences in the complexity behavior of SESD and
of relaxed SESD.

\subsection{Low-SNR Regime}
\label{lowsnrcomplexity}

\subsubsection*{Finite-lattice case}

SESD visits a node~$\bmx^{(i)}$ if the SC $d\big(\bmx^{(i)}\big) \leq
r$ is satisfied. Since the minimum search radius of SESD (guaranteeing
that the corresponding hypersphere contains at least one lattice
point) is given by the Euclidean distance between the received vector
and the ML solution, we can conclude that SESD visits at least all
nodes~$\bmx^{(i)}$ satisfying
\begin{align} \label{eq:sufficientconditionfiniteSESD}
  d\big(\bmx^{(i)}\big) \leq
  \min_{\bmx\in\setX^\MT}\|\bmrtilde-\bR\bmx\| = r_\mathrm{min}.
\end{align}
Thanks to radius reduction, the final (with respect to the repeated
tree-traversals as described above) search radius of SESD corresponds
to~$r_\mathrm{min}$. We can therefore conclude that the genie-aided
choice \mbox{$r=r_\mathrm{min}$} for the search radius initialization,
indeed, yields a lower bound on the number of nodes visited by SESD.
Denoting the transmitted data vector as~$\bmx'\in\setX^\MT$, we obtain
a lower bound on~$r_\mathrm{min}$ as follows:
\begin{align}
  r_\mathrm{min} & \stackrel{a)}{=} \min_{\bmx\in\setX^\MT}
  \big\|\bmntilde+\bR(\bmx'-\bmx)\big\| \geq \nonumber \\
  & \stackrel{\mathrm{b)}}{\geq} \min_{\bmx\in\setX^\MT} \Big(
  \|\bmntilde\|-\big\|\bR(\bmx-\bmx')\big\| \Big)= \nonumber \\
  & = \|\bmntilde\|-
  \max_{\bmx\in\setX^\MT}\big\|\bR(\bmx-\bmx')\big\| \geq \nonumber \\
  & \stackrel{\mathrm{c)}}{\geq}
  \|\bmntilde\|-\sigma_\mathrm{max}(\bR)K
  \label{eq:SESDcomplxlowerbound}
\end{align}
where a) results from~$\bmrtilde = \bR\bmx' + \bmntilde$, b) follows
from the inverse triangle inequality, and c) is a consequence of the
Rayleigh-Ritz theorem~\cite{HJ85} ($\sigma_\mathrm{max}(\bR)$ denotes
the largest singular value of $\bR$) and the fact that the
constant~\mbox{$K=\max_{\bmx,\bmx'\in\setX^\MT}\|\bmx-\bmx'\|$} is
finite in the finite-lattice case.
From~\eqref{eq:SESDcomplxlowerbound} we can infer that SESD visits at
least all nodes~$\bmx^{(i)}$ satisfying
\begin{align} \label{eq:sufficientconditionfiniteSESDLB}
  d\big(\bmx^{(i)}\big) \leq \|\bmntilde\|-\sigma_\mathrm{max}(\bR)K.
\end{align}
Note that setting the search radius equal to the right-hand side (RHS)
of~\eqref{eq:sufficientconditionfiniteSESDLB} will no longer guarantee
that the corresponding hypersphere contains at least one lattice
point.  This is, however, irrelevant for the point we want to make
next.

Defining~\mbox{$\bmv=\bmntilde/\sqrt{N_o}$}, we can
rewrite~\eqref{eq:sufficientconditionfiniteSESDLB} as
\begin{align}  \label{eq:sufficientconditionfiniteSESDLBnormalized}
  \Bigg\|\bmv^{(i)} +
  \frac{\bR_i\big(\bmx'^{(i)}-\bmx^{(i)}\big)}{\sqrt{N_o}}\Bigg\| \leq
  \|\bmv\|-\frac{\sigma_\mathrm{max}(\bR)K}{\sqrt{N_o}}.
\end{align}
We can now see that for~$N_o\to\infty$ and hence for~$\SNR\to0$, the
condition in~\eqref{eq:sufficientconditionfiniteSESDLBnormalized}
reduces to~$\big\|\bmv^{(i)}\big\|\leq\|\bmv\|$, which is trivially
satisfied for~$i>1$ for every realization of~$\bmntilde$ and for
given~$\bR$. In this case, SESD visits at least all nodes down to and
including the level just above the leaf level.  An intuitive
explanation for this effect is that the noise can shift the transmit
vector arbitrarily far away from the finite lattice and hence, the RHS
of~\eqref{eq:SESDcomplxlowerbound} and consequently~$r_\mathrm{min}$
can become arbitrarily large, which inhibits efficient tree pruning.

\subsubsection*{Infinite-lattice case}

The fundamental difference between the complexity of SESD and of relaxed
SESD is due to the fact that in the infinite-lattice case, the maximum
distance between the received vector~$\bmr$ and the nearest lattice
point is bounded for a given (full rank)~$\bR$. The complexity of
relaxed SESD can therefore be lower than that of SESD.  The downside
is that relaxed SESD is not guaranteed to find the ML solution, which
necessarily has to be a point in the finite lattice~$\Lattice(\bG)$.
Specifically, if relaxed SESD delivers a lattice point
outside~$\Lattice(\bG)$, this point needs to be remapped onto the
finite lattice.  Corresponding remapping approaches are described
in~\secref{relaxedMLDperformance}.  In the following, we derive an
analytic upper bound on the complexity of relaxed SESD.

The number of nodes~$\bmxREL^{(i)}$ on tree level~$i$ that satisfy the
SC with radius~$r$ is denoted by~$\CREL_{i}(r)$ and corresponds to the
number of lattice points in~$\LatticeREL(\bR_i)$ that are within a
$2(\MT-i+1)$-dimensional (real-valued) hypersphere of radius~$r$
centered at~$\bmr^{(i)}$. This number can be upper-bounded by dividing
the volume of a $2(\MT-i+1)$-dimensional hypersphere of
radius~$r+\mu_i$ by the volume of a Voronoi cell
in~$\LatticeREL(\bR_i)$~\cite{BK98}, where~$\mu_i$ corresponds to the
covering radius of the lattice~$\LatticeREL(\bR_i)$~\cite{CS99}.  More
specifically, we have~\cite{Wills91}
\begin{align} \label{eq:complexityupperboundunbestimmt} 
  \CREL_{i}(r) \leq \frac{V_{(\MT-i+1)}(r+\mu_i)}{\mathrm{Vol}\big(\LatticeREL(\bR_i)\big)}
\end{align}
where~$V_k(r) = \frac{\pi^k}{k!}r^{2k}$ denotes the volume of
a~$2k$-dimensional hypersphere with radius~$r$ and the volume of a
Voronoi cell of~$\LatticeREL(\bR_i)$ is
\mbox{$\mathrm{Vol}\big(\LatticeREL(\bR_i)\big)=\det(\bR_i^H\bR_i)$}.

In order to derive an upper bound on the total complexity
\mbox{$\CREL=\sum_{i=1}^{\MT}\CREL_i(r)$} of relaxed SESD, we consider
the case where only the first radius update is performed. The first
leaf node found by relaxed SESD corresponds to the Babai
point~$\bmxhatREL^\BABAI$~\cite{BABAI86} and hence the radius after
the first update from the initial value~$r=\infty$ is given
by~$d\big(\bmxhatREL^\BABAI\big)$, which can be upper-bounded
as~\cite{BABAI86,AEVZ02}
\begin{align} \label{eq:babaibound} 
  d\big(\bmxhatREL^\BABAI\big) = \big\|\bmrtilde-\bR\bmxhatREL^\BABAI\big\| \leq
  \sqrt{\frac{1}{2}\sum_{j=1}^{\MT}R^2_{j,j}} = \beta.
\end{align}
Since the covering radius~$\mu_1$ is also upper-bounded by~$\beta$
\cite{Kannan87,BK98} and, more generally, $\mu_i$ can be upper-bounded
as~$\mu_i\leq\sqrt{\frac{1}{2}\sum_{j=i}^{\MT}R^2_{j,j}}=\gamma_i$, we
have $r+\mu_i\leq\beta+\gamma_i$, which yields an upper bound on the
total complexity of relaxed SESD according to
\begin{align} \label{eq:totalcomplexityupperbound} 
  \CREL \leq
  \sum_{i=1}^{\MT}
  \frac{V_{(\MT-i+1)}(\beta+\gamma_i)}{\det(\bR_i^H\bR_i)}.
\end{align}

\subsection{High-SNR Regime}
\label{highSNRcomplexity}

In the high-SNR regime, relaxed SESD is likely to return the ML
solution.  Moreover, if~$\|\bmntilde\|$ is sufficiently small, relaxed
SESD and SESD find the same leaf nodes, which will belong to the
finite lattice~$\Lattice(\bG)$. Consequently, relaxed SESD and SESD
will operate with the same search radii and hence the same SCs.
However, the two detectors will, in general, not exhibit the same
complexity, as the SESD additionally (to the SC) takes into account
the finite alphabet nature of~$\setX\subset\CZ$, whereas relaxed SESD
visits \emph{all} points in~$\CZ$ satisfying the SC.  Therefore, in
the high-SNR regime, SESD tends to result in smaller complexity than
relaxed SESD.

\subsection{Numerical Complexity Assessment}
\label{numericalcomplexityassessment}

A simple numerical result serves to demonstrate the potential
complexity savings of relaxed SESD over SESD in the low-SNR regime. We assume a MIMO channel
with~$\bR=\bI_\MT$ and $\MT=4$. Counting all the nodes down to
and including the level just above the leaf level and noting that only
one leaf node is visited by SESD if~$\bR$ is a diagonal matrix, leads
to the corresponding complexity
\begin{align*}
  C = \frac{\abs{\setX}^\MT-1}{\abs{\setX}-1}
\end{align*}
of SESD for~$\SNR\to0$ (see the discussion in the last paragraph of
\secref{lowsnrcomplexity}). For a 16-QAM alphabet, this leads to
\mbox{$C=4369$} nodes. The total complexity of relaxed SESD is
upper-bounded by (16), which can be refined by noting that relaxed SESD
visits only one leaf node in the case of~\mbox{$\bR=\bI_\MT$}. Consequently, we have
\begin{align*}
  \CREL \leq 1+ \sum_{i=2}^{\MT}
  V_{(\MT-i+1)}\Bigg(\!\sqrt{\frac{\MT}{2}}+\sqrt{\frac{\MT-i+1}{2}}\Bigg)
\end{align*}
which shows that at most~1928 nodes will be visited, irrespectively of
the SNR. We can therefore conclude that relaxation can result in a
significant complexity reduction at low SNR values.

\figref{fig:CPXremapping} shows the complexity\footnote{All simulation
  results are for an uncoded \mbox{$\MT=4$} MIMO system with 16-QAM
  symbol constellation using Gray labeling. The entries of~$\bH$ are
  i.i.d.\ circularly symmetric complex Gaussian distributed with unit
  variance.  For (relaxed) SESD, we use sorted QR-decomposition (SQRD)
  according to~\cite{wuebben01}, for LR, a complex-valued variant of
  the LLL algorithm with~$\delta=3/4$ and SQRD preprocessing as
  described in~\cite{GW05,WBKK04}. All complexity and performance
  simulations are averaged over 640,000 channel realizations and for
  each channel realization a single noise realization has been
  generated.}  corresponding to SESD (which yields ML performance) and
to relaxed SESD (denoted by R-SESD). Consistent with the observations
made in Sections~\ref{lowsnrcomplexity} and~\ref{highSNRcomplexity},
we see that relaxation leads to complexity savings in the low-SNR
regime, but results in higher complexity in the high-SNR regime. We
can furthermore see that relaxation followed by LR (denoted by
LR-SESD) leads to lower complexity than SESD for all SNR values.

The complexity reductions of R-SESD and LR-SESD over SESD come,
however, at a significant performance loss in terms of error-rate and
in the case of LR-SESD also at increased computational complexity in
the preprocessing stage caused by the need for LR (realized through
the LLL algorithm~\cite{LLL82}, for example). For naive lattice
decoding, i.e., estimates~$\bmuhatREL^\ML\notin \Lattice(\bG)$ are
simply discarded and an error is declared,~\figref{fig:BERremapping}
shows a 3\,dB~SNR loss of LR-SESD compared to SESD (MLD). We emphasize
that the performance of R-SESD, not shown
in~\figref{fig:BERremapping}, is identical to that of LR-SESD; the
associated complexities are, however, different
(cf.~\figref{fig:CPXremapping}). Note that naive lattice decoding was
shown to achieve full diversity, while leading to an unbounded SNR gap
(growing logarithmically in SNR) for MIMO systems with an equal number
of transmit and receive antennas~\cite{TK07}. In the next section, we
discuss several approaches to the mitigation of this performance loss,
i.e., we consider algorithms for remapping
estimates~$\bmuhatREL^\ML\notin \Lattice(\bG)$ onto~$\Lattice(\bG)$.


\section{Remapping Methods}
\label{relaxedMLDperformance}

If the estimate delivered by relaxed SESD satisfies
\mbox{$\bmuhatREL^\ML \in \Lattice(\bG)$}, the lattice
point~$\bmuhatREL^\ML$ corresponds to the ML
solution~\eqref{eq:constrainedMLrule},
i.e.,~$\bmuhatREL^\ML=\bmuhat^\ML$. If~$\bmuhatREL^\ML \notin
\Lattice(\bG)$ and the performance of naive lattice decoding is not
sufficient, the solution delivered by relaxed SESD needs to be
remapped onto the finite lattice~$\Lattice(\bG)$. Since all
constellation points~$\bmx\in\setX^\MT$ (corresponding to
elements~$\bmu\in\Lattice(\bG)$ through the relation\footnote{In the
  remainder of the paper, we treat vectors~$\bmxREL\in(\CZ)^\MT$ and
  \mbox{$\bmuREL\in\LatticeREL(\bG)$} as well as~$\bmx\in\setX^\MT$
  and~$\bmu\in\Lattice(\bG)$ as interchangeable as they are related
  through the one-to-one mappings $\bmuREL=\bG\bmxREL$ and
  $\bmu=\bG\bmx$, respectively. Of course, this assumes that~$\bG$ has
  full rank.} $\bmu=\bG\bmx$) satisfy~\eqref{eq:formalsetdefinition},
determining whether~$\bmuhatREL^\ML$ is in the finite
lattice~$\Lattice(\bG)$ can be performed efficiently through
element-wise boundary checks of the real and imaginary parts
of~$\bmxhatREL^\ML$.  In the following, we briefly review an existing
remapping method known as quantization and describe novel remapping
approaches that realize different performance/complexity trade-offs.

\subsection{Quantization}

The standard remapping approach known in the literature is
quantization and amounts to remapping
\mbox{$\bmxhatREL^\ML\notin\setX^\MT$} to the closest (in Euclidean
distance) symbol vector~$\bmxhat_\mathrm{q} \in \setX^\MT$ according
to~\cite{WF03,WBKK04}
\begin{align} \label{eq:quantization}
  \bmxhat_\mathrm{q} =
  \argmin_{\bmx\in\setX^\MT} \big\|\bmxhatREL^\ML-\bmx\big\|.
\end{align}
As quantization corresponds to element-wise slicing of the entries
of~$\bmxhatREL^\ML$ to the nearest candidates in~$\setX$, it requires
low computational complexity. The resulting
estimate~$\bmxhat_\mathrm{q}$ does, however, not necessarily
correspond to the ML solution~\eqref{eq:constrainedMLrule} and hence
leads to a (potentially significant) performance loss.

\figref{fig:BERremapping} shows the performance obtained by LR-aided
SESD followed by remapping through quantization (denoted by LR-SESD,
quant.).  The numerical results indicate a slight improvement over
naive lattice decoding, but still show a 2.5\,dB SNR loss compared to
ML performance. Finally, \figref{fig:BERremapping} shows the
performance obtained by LR-aided successive interference cancellation
(SIC) on the infinite lattice with remapping through naive lattice
decoding (denoted by LR-SIC, naive). The resulting performance loss,
compared to relaxed SESD with remapping through naive lattice
decoding, is low (approximately 0.25\,dB SNR), whereas the complexity
of LR-SIC is significantly lower than that of LR-SESD.

\subsection{Optimal Lattice Remapping} \label{optimumlatticeremapping}

The discussion above brings out the importance of remapping.  It is
therefore natural to ask for the optimum lattice remapping strategy in
the sense of maximizing the probability of the result of remapping to
correspond to the transmitted symbol vector given the result of the
relaxed detector~$\bmuhatREL$. Let~$\bmu'\in\Lattice(\bG)$ be the
transmitted vector and assume that the relaxed detector yields
\begin{align*}
  \DREL(\bmr)=\DREL(\bmu'+\bmn)=\bmuhatREL, \quad
  \bmuhatREL\in\LatticeREL(\bG)\!\setminus\! \Lattice(\bG).
\end{align*}
Optimal lattice remapping according to the criterion described above
now amounts to identifying the lattice point $\bmu\in\Lattice(\bG)$
which maximizes $\Pr{\bmu'= \bmu|\DREL(\bmr)=\bmuhatREL}$, i.e.,
\begin{align*}
  \bmuhat_{\mathrm{opt}} = \argmax_{\bmu\in\Lattice(\bG)}
  \Pr{\bmu'=\bmu|\DREL(\bmr)=\bmuhatREL}\!.
\end{align*}
We emphasize that the optimum remapping rule depends on~$\DREL(\bmr)$
and hence, on the relaxed MIMO detector employed. Using Bayes's
theorem and assuming that all transmit vectors are equally likely, we
get
\begin{align} \label{eq:remapperoptimum} \bmuhat_{\mathrm{opt}} =
  \argmax_{\bmu\in\Lattice(\bG)} \Pr{\DREL(\bmr)=\bmuhatREL|\bmu'=\bmu}
\end{align}
which amounts to solving a finite-lattice MLD
problem~\eqref{eq:constrainedMLrule} based on the relaxed
estimate~$\bmuhatREL$ instead of the received vector~$\bmr$. We stress
that relaxed MLD followed by optimum lattice remapping according
to~\eqref{eq:remapperoptimum} will, in general, not achieve ML
performance as remapping is conducted based on~$\bmuhatREL$ rather
than on~$\bmr$. For the sake of simplicity of exposition, in the
following, we write $\Pr{\DREL(\bmr)=\bmuhatREL|\bmu}$ instead of
$\Pr{\DREL(\bmr)=\bmuhatREL|\bmu'=\bmu}$.

\subsubsection*{Closest-vector remapping}

The optimal remapping rule in~\eqref{eq:remapperoptimum} can, in
general, not be expressed in closed form, as it depends on the
decision region of the relaxed MIMO detector. We can, however, obtain
a simple remapping strategy that achieves near-optimal performance by
computing an approximation of~\eqref{eq:remapperoptimum} as follows.
We start by noting that
\begin{align} \label{eq:latticeprobability}
  \Pr{\DREL(\bmr)=\bmuhatREL|\bmu} =
  \int_{\setD(\bmuhatREL)} \!
  f_G( \bmw - \bmu ) \,\mathrm{d}\bmw
\end{align}
where 
\begin{align*}
f_G(\bmw) = (\pi N_o)^{-\MR}\exp\big(\!-N_o^{-1}\|\bmw\|^2\big)
\end{align*}
is the joint probability density function of a multi-variate complex
Gaussian with i.i.d.\ circularly symmetric components each of which has
variance~$N_o$ and
\begin{align*}
\setD(\bmuhatREL)
  = \Big\{ \bmw \in \C^\MT \mid \DREL(\bmw) = \bmuhatREL\Big\}
\end{align*}
denotes the decision region of the relaxed MIMO detector employed.
Next, we consider optimal remapping for relaxed MLD, i.e.,
$\DREL(\bmr)=\DREL^\ML(\bmr)$ as defined
in~\eqref{eq:unconstrainedMLrule} and $\bmuhatREL=\bmuhatREL^\ML$, and
note that in this case~\eqref{eq:latticeprobability} formally
corresponds to the probability of mistakenly decoding the transmitted
point~$\bmu$ for~$\bmuhatREL^\ML$.  This probability can be
upper-bounded as
\begin{align} \label{eq:Prupperbound}
  \Pr{\DREL^\ML(\bmr)=\bmuhatREL^\ML\big|\bmu}  \leq  
  Q\Bigg(\frac{\big\| \bmuhatREL^\ML -
    \bmu\big\|}{\sqrt{2N_o}}\Bigg)
\end{align}
where~$Q(a)=(2\pi)^{-\frac{1}{2}}\!\int_{a}^\infty\exp(-\frac{x^2}{2})\,\mathrm{d}x$.
Replacing the function to be maximized over
in~\eqref{eq:remapperoptimum} by the upper bound
in~\eqref{eq:Prupperbound} leads to a novel remapping rule, which we
refer to as closest-vector remapping (CVR) and which is given by
\begin{align} \label{eq:cvremapper} \bmuhat_{\mathrm{cv}} =
  \argmin_{\bmu\in\Lattice(\bG)}\big\|\bmuhatREL^\ML-\bmu\big\|.
\end{align}
The solution of~\eqref{eq:cvremapper} corresponds to the
point~$\bmuhat_{\mathrm{cv}}\in\Lattice(\bG)$ that is closest (in
Euclidean distance) to the lattice
point~$\bmuhatREL^\ML\in\LatticeREL(\bG)\!\setminus\! \Lattice(\bG)$.
We emphasize that CVR amounts to solving a finite-lattice CVP and can,
hence, be carried out by (non-relaxed) SESD with the relaxed
ML-estimate~$\bmuhatREL^\ML$ taking the role of the received point.
Note that CVR according to~\eqref{eq:cvremapper} exhibits structural
similarities to quantization according to~\eqref{eq:quantization}.
Formally, quantization can be seen as CVR with~$\bG=\bI_\MT$, which
amounts to ignoring the structure of the lattice generated by~$\bG$.

\subsubsection*{Numerical performance and complexity results}

\begin{figure}[tb]
  \centering
  \includegraphics[width=\figscale\columnwidth]{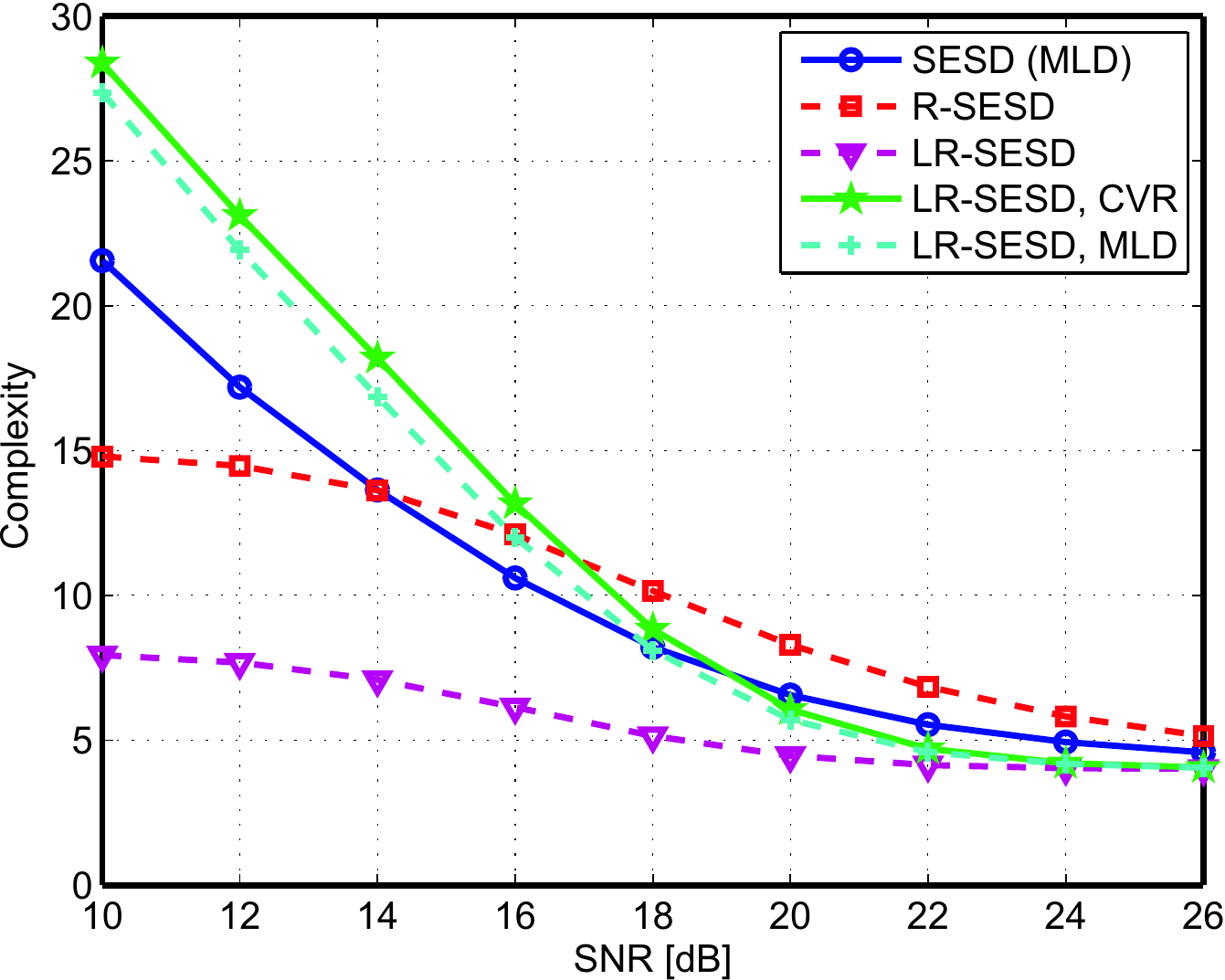}
  \figspacepre
  \caption{Complexity of SESD (MLD), relaxed SESD (R-SESD), and
    LR-aided SESD (LR-SESD).}
  \figspacepost
  \label{fig:CPXremapping}
\end{figure}

\begin{figure}[tb]
  \centering
  \includegraphics[width=\figscale\columnwidth]{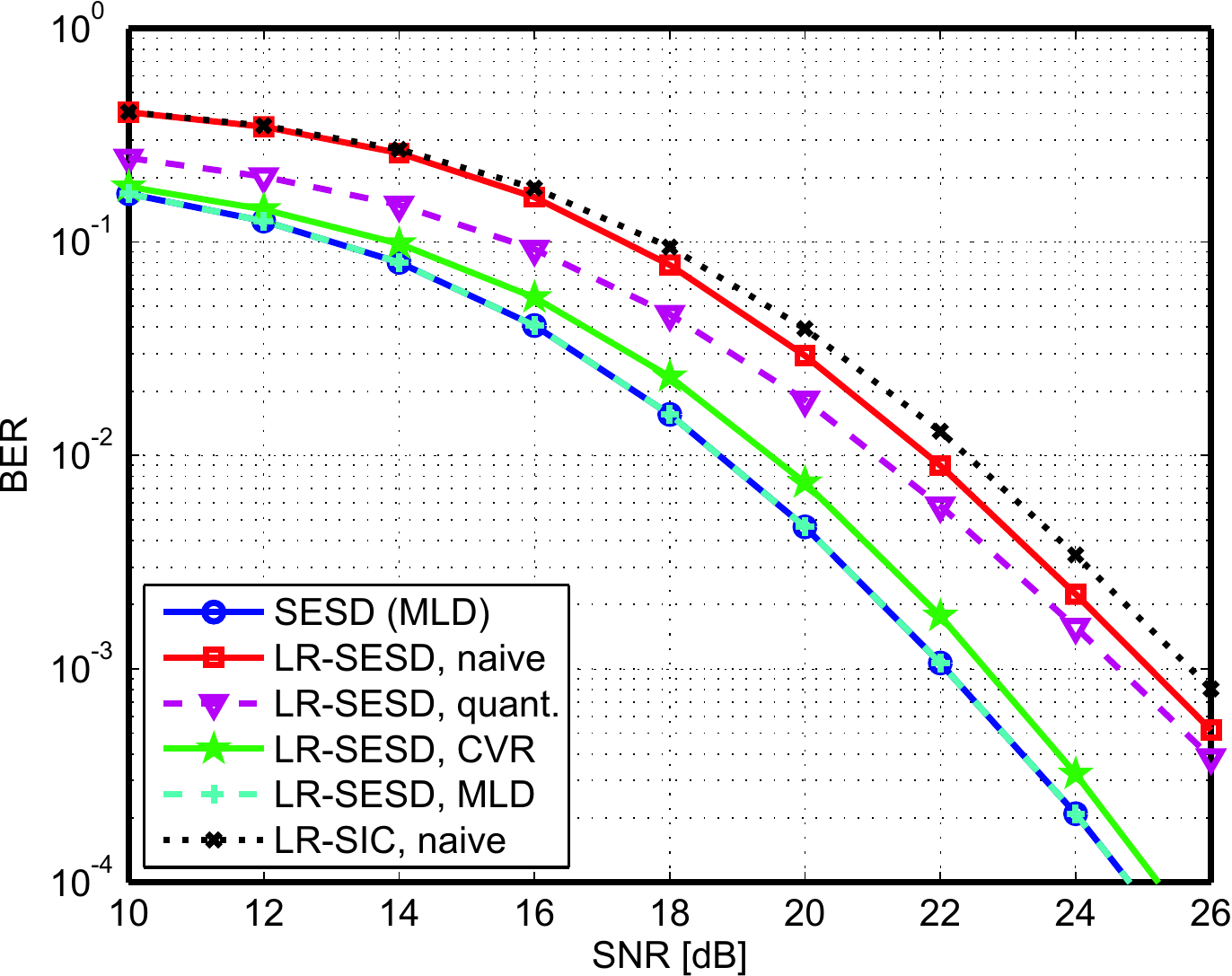}
  \figspacepre
  \caption{Bit error rate (BER) performance comparison of SESD (MLD), LR-SESD, and LR-SIC.}
  \figspacepost
  \label{fig:BERremapping}
\end{figure}

\figref{fig:BERremapping} compares the performance of relaxed SESD
followed by CVR to that obtained by relaxed SESD followed by
quantization. We observe that CVR significantly outperforms
quantization (and naive lattice decoding) and yields close-to-ML
performance.  This improvement comes, however, at the cost of having
to solve a finite-lattice CVP, namely~\eqref{eq:cvremapper}, in
contrast to simple element-wise slicing in the case of quantization
and to simply discarding estimates that do not belong
to~$\Lattice(\bG)$ in the case of naive lattice decoding, both of
which exhibit significantly smaller complexity.

We next compare the overall complexity (i.e., the total complexity of
both SESD runs) of LR-aided SESD followed by CVR (denoted by LR-SESD,
CVR) to that of SESD operating directly on the finite lattice, i.e.,
SESD (MLD).  \figref{fig:CPXremapping} shows that LR-SESD with CVR
outperforms SESD (MLD) \emph{only} in the very-high-SNR regime with
the corresponding complexity savings being minor. In addition, this
comparison favors \mbox{LR-SESD} with CVR as the additional complexity associated
with LR has been neglected.

\subsection{Two-Stage Detection}

The discussion above leads us to the question of whether
using~$\DREL(\bmr)=\bmuhatREL$ for remapping instead of~$\bmr$ is
sensible. After all, by mapping~$\bmr$ onto~$\bmuhatREL$ we are
potentially discarding useful information. In the following, we show
that, indeed, performing remapping based on~$\bmr$ rather than
on~$\bmuhatREL$ leads to a performance improvement (compared to, e.g.,
CVR) without requiring additional complexity. 

To this end, we propose a new detector structure, which consists of
two stages. The first stage corresponds to relaxation
from~$\Lattice(\bG)$ to~$\LatticeREL(\bG)$ (possibly followed by LR)
and using a relaxed MIMO detector. The second stage is invoked
\emph{only} if the first stage yields a result that does \emph{not}
belong to the finite lattice~$\Lattice(\bG)$ and corresponds to a MIMO
detector operating on the finite lattice.  This second-stage detector
\emph{uses the received vector}~$\bmr$ instead of~$\bmuhatREL$. We
emphasize that if both stages employ MLD, the two-stage detector
achieves ML performance, since in the case of
\mbox{$\bmuhatREL^\ML\notin\Lattice(\bG)$}, the second stage performs
non-relaxed MLD on the basis of~$\bmr$.

\subsubsection*{Numerical performance and complexity results}

\figref{fig:CPXremapping} shows the overall complexity of two-stage
detection (i.e., the total complexity across the two stages) using
LR-aided SESD in the first stage and SESD in the second stage (denoted
by LR-SESD, MLD). Again, the complexity associated with LR is ignored.
Consulting the corresponding performance results
in~\figref{fig:BERremapping}, we can conclude that two-stage detection
employing SESD in both stages and LR in the first stage outperforms
\mbox{LR-SESD} with CVR both in terms of complexity \emph{and}
performance and, hence, the performance loss associated with
\mbox{LR-SESD} followed by CVR (compared to ML performance) must
result from the fact that~$\bmuhatREL^\ML$ rather than~$\bmr$ is used
for remapping. We can furthermore conclude that in the high-SNR regime
the two-stage approach exhibits slightly lower complexity than SESD,
while realizing the same (i.e., ML) performance.  We finally note that
the computational complexity of LR and the complexity of checking
whether~$\bmuhatREL\in\Lattice(\bG)$ has been neglected in this
comparison.

 
\vspace{-0.3cm}

\section{Conclusions}

We demonstrated that relaxation from finite to infinite lattices
combined with lattice reduction (LR) can reduce the (tree-search)
complexity of Schnorr-Euchner sphere decoding (SESD), but generally
leads to a significant performance loss. This performance loss is
caused by the fact that the relaxed detector does not necessarily
deliver estimates that belong to the finite lattice and hence does not
realize ML performance. Remapping methods that yield (close-to) ML
performance necessitate solving finite-lattice closest vector problems
(CVPs).  Hence, if ML performance is required, using SESD directly on
the finite lattice seems to require smaller complexity, as remapping
is avoided and only one (finite-lattice) CVP needs to be solved. We
proposed an ML-optimal two-stage detector which was shown to require
lower complexity than SESD on the finite lattice provided i) the SNR
is high enough and ii) the complexity corresponding to LR and to
checking whether estimates correspond to the finite lattice is
neglected.  Generally speaking, we conclude that relaxation and LR are
better suited for use in conjunction with low-complexity detectors
(such as linear detection or SIC) with overall sub-optimal
performance.

\bibliographystyle{IEEEtran} \bibliography{IEEEabrv,08asilomar_lll}

\end{document}